# Short Circuit Synchronized Electric Charge Extraction (SC-SECE): a tunable interface for wideband vibration energy harvesting

Adrien MOREL[1,2*], Adrien BADEL[2], Pierre GASNIER[1], David GIBUS[1,2], Gaël PILLONNET[1]

[1] Univ. Grenoble Alpes, CEA, LETI, MINATEC, F-38000 Grenoble, France
[2] Univ. Savoie Mont Blanc, SYMME, F-74000 Annecy, France

*adrien.morel@cea.fr

*Abstract*— In this paper, we present a new harvesting interface, called Short Circuit Synchronous Electric Charge Extraction (SC-SECE). The SC-SECE strategy includes a tunable short-circuit time thanks to two tuning parameters, $\phi_S$ and $\Delta\phi$. $\phi_S$ stands for the phase between the mechanical displacement extrema and the energy harvesting event. $\Delta\phi$, stands for the angular time spent in the short-circuit phase. The theoretical analysis and modelling of this short-circuit influences are derived in this paper. When associated with highly coupled harvesters, it is shown that both the harvested power and bandwidth are greatly improved. These results have been numerically validated and they demonstrate the potential of this strategy for extending the bandwidth of piezoelectric vibration energy harvesters.

## I. INTRODUCTION

In order to make small systems and sensors autonomous scavenging ambient energy has been widely investigated in the last two decades as an alternative to batteries. Piezoelectric energy harvesters (PEH) are of particular interest in closed confined environments, where there are few solar radiations and thermal gradients.

In order to maximize the energy harvested from piezoelectric harvesters, the electrical interface is a key point to consider. Several non-linear synchronous strategies, such as Synchronous Electric Charge Extraction (SECE) and Synchronized Switch Harvesting on Inductance (SSHI) have been introduced [1], and have been implemented using discrete components [2,3] or dedicated ASIC [4,5]. Those strategies exhibit high performance for lowly coupled and/or highly damped piezoelectric harvesters. However, for highly coupled and/or lowly damped piezoelectric harvesters, these strategies may overdamp the mechanical resonator, leading to low performances. To face this challenge, researchers started to propose new tunable strategies inducing lower damping [6], and even tuning the PEH resonant frequency thanks to the important influences of the electrical interface on the mechanical resonator [7].

In this paper, we introduce a strategy based on the SECE interface, which introduces a tunable short-circuit phase. As detailed extensively in this paper, this short-circuit allows to reduce the damping induced by the electrical interface. Furthermore, for high coupling harvesters, it allows to tune the PEH resonant frequency, leading to an enhanced harvesting bandwidth.

## II. THEORETICAL MODELLING

### A. Linear PEH modelling

A linear PEH under a periodic excitation can usually be modelled by a system of linear differential equations, given by (1).

$$\begin{cases} M\ddot{x} + D\dot{x} + K_{sc}x + \alpha v_p = -F = -M\ddot{y} \\ i = \alpha\dot{x} - C_p\dot{v}_p \\ x(t) = X_m\cos(\theta) = X_m\cos(\omega t) \end{cases} \quad (1)$$

where $y$, $F$ and $x$ stand for the ambient displacement, the force applied on the PEH, and the tip mass displacement, respectively. $M, K_{sc}, D, C_p$ and $\alpha$ stand for the equivalent mass of the PEH, its short-circuited stiffness, its mechanical damping, the capacitance of the piezoelectric material, and the piezoelectric force coefficient, respectively. Fig.1 shows the electrical circuit modelling these equations.

### B. Expression of the piezoelectric voltage

In order to find the harvested power expression, we have to solve (1) and find the mechanical displacement magnitude $X_m$. As $v_p$ is a variable in (1), finding a linear expression of $v_p$ would greatly simplify the calculations. The piezoelectric harvester is either working in open or short-circuit, as illustrated in Fig.2. Thus, the piezoelectric voltage during a single vibration period is given by (2).

$$v_p(\theta) = \begin{cases} \dfrac{\alpha}{C_p}\int_{\phi_s+\Delta\phi-\pi}^{\theta} \dot{x}(\theta)d\theta, \forall\theta \in\,]\phi_s+\Delta\phi-\pi, \phi_s] \\ 0, \forall\theta \in\,]\phi_s, \phi_s+\Delta\phi] \\ \dfrac{\alpha}{C_p}\int_{\phi_s+\Delta\phi}^{\theta} \dot{x}(\theta)d\theta, \forall\theta \in\,]\phi_s+\Delta\phi, \phi_s+\pi] \\ 0, \forall\theta \in\,]\phi_s+\pi, \phi_s+\Delta\phi+\pi] \end{cases} \quad (2)$$

$\phi_s \in [0,\pi]$ corresponds to the angular phase between the harvesting process and the precedent displacement extremum. $\Delta\phi \in [0,\pi]$ stands for the angular time spent in the short-circuit phase. A system implementing this extraction strategy is depicted in Fig.1, and an example of the voltage waveform is shown in Fig.2. As expressed by (2), $v_p$ is not sinusoidal. Only the first harmonic of $v_p$ is considered in order to analytically solve (1) while simplifying the calculations.

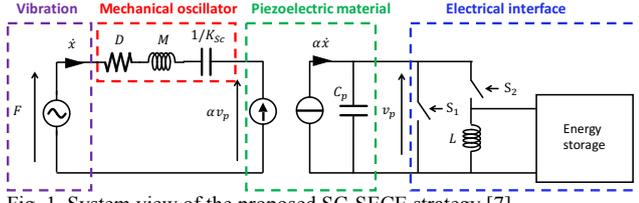

Fig. 1. System view of the proposed SC-SECE strategy [7]

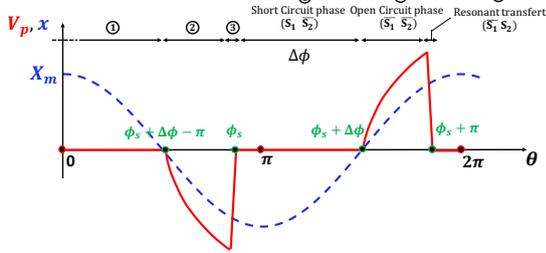

Fig. 2. Displacement (blue) and voltage (red) waveforms of the SC-SECE

From (2), we eventually find the expression of the Fourier series coefficients $a_1$ and $b_1$ associated with $v_p$:

$$\begin{cases} a_1 = \frac{\alpha X_m}{\pi C_p}[\pi - \Delta\phi + \frac{\sin(2\phi_s + 2\Delta\phi)}{2} + \frac{\sin(2\phi_s)}{2} \\ \qquad\qquad + 2\cos(\phi_s + \Delta\phi)\sin(\phi_s)] \\ b_1 = -\frac{\alpha X_m}{\pi C_p}[\cos\phi_s + \cos(\phi_s + \Delta\phi)]^2 \end{cases} \quad (3)$$

Hence, the first harmonic of $v_p$, $\underline{v_{p_1}}$, is expressed by (4).

$$\underline{v_{p_1}} = \underline{x}\left(\frac{a_1}{X_m} - j\frac{b_1}{X_m}\right) = \underline{x}(a_1^* - jb_1^*) \quad (4)$$

where $\underline{x}$ is the mechanical displacement in the Fourier domain. $a_1^*$ and $b_1^*$ are the first Fourier coefficients $a_1$ and $b_1$ divided by the displacement amplitude, $X_m$.

*C. Expression of the harvested power*

Due to the filtering effect of the resonator, we consider that the first voltage harmonic may impact the PEH dynamics. We can hence substitute the voltage expression (4) in (1) to find the displacement amplitude. Solving (1) in the Fourier domain, the mechanical displacement amplitude can be expressed by (5).

$$X_m = \frac{|M\ddot{y}|}{\sqrt{(K_{sc} - M\omega^2 + \alpha a_1^*)^2 + (\omega D + \alpha b_1^*)^2}} \quad (5)$$

For each vibration's semi-period, the harvested energy is the one stored in $C_p$ when $\theta = \phi_s$. Thus, from (1), we can derive the harvested power expression:

$$P_{harv} = \frac{\omega\alpha^2}{2\pi C_p}X_m^2[\cos(\phi_s + \Delta\phi) + \cos(\phi_s)]^2 \quad (6)$$

This power $P_{harv}$, divided by the maximum harvestable power $P_{max} = D^{-1}|M\ddot{y}|^2/8$ [6] has been computed in Figure 3 with optimized parameters $(\phi_s, \Delta\phi)$ as a function of the normalized vibration frequency $\Omega_m = \omega/\omega_0$ with three normalized squared coupling coefficients $k_m^2$ and a mechanical quality factor of $Q_m=25$. These normalized parameters have been extensively described in [6,7]. We can observe that the bandwidth gain becomes more important as $k_m^2$ is increased.

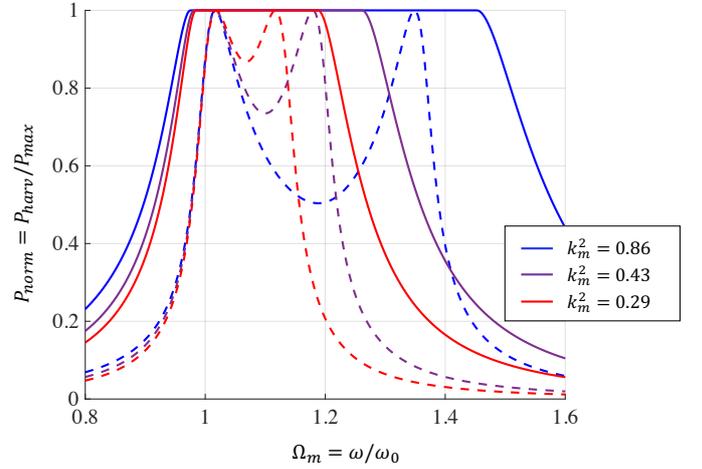

Fig. 3. Frequency responses of highly coupled piezoelectric harvesters with optimized resistive loads (dashed lines) and the proposed SC-SECE interface (straight lines)

III. CONCLUSION

In this paper, we present the SC-SECE strategy, which is based on a combination of the SECE interface and a tunable short-circuit. Numerical results show that this strategy could be used to both enhance the harvested power and the harvesting bandwidth of highly coupled PEH Experiments confirming these theoretical predictions are currently being run, and will be presented during the conference.

REFERENCES

[1] E. Lefeuvre *et al.*, "A comparison between several vibration-powered piezoelectric generators for standalone systems", Sensors and Actuators A: Physical, vol. 126, no. 2, pp. 405–416, Feb. 2006.

[2] G. Shi *et al.*, "An efficient self powered synchronous electric charge extraction interface circuit for piezoelectric energy harvesting systems" Journal of Intelligent Material Systems and Structures, vol. 27, no. 16, pp. 2160–2178, Sep. 2016.

[3] Y. Wu *et al.*, "Piezoelectric vibration energy harvesting by optimized synchronous electric charge extraction" Journal of Intelligent Material Systems and Structures 24(12): 1445–1458, 2012.

[4] A. Quelen *et al.*, "A 30nA Quiescent 80nW to 14mW Power Range Shock-Optimized SECE-based Piezoelectric Harvesting Interface with 420% Harvested Energy Improvement", IEEE International Solid State Circuit Conference, 2018.

[5] T. Hehn *et al.*, "A Fully Autonomous Integrated Interface Circuit for Piezoelectric Harvesters", IEEE Journal of Solid-State Circuits, vol. 47, no. 9, pp. 2185–2198, Sep. 2012.

[6] A. Morel *et al.*, "Regenerative synchronous electrical charge extraction for highly coupled piezoelectric generators", *IEEE Midwest symposium of circuits and systems (MWSCAS) 2017,* 2017.

[7] A. Morel *et al.*, "Short Circuit Synchronous Electric Charge Extraction(SC-SECE) Strategy for Wideband Vibration Energy Harvesting", *IEEE International Symposium of Circuits And Systems (ISCAS) 2018*, 2018.